\documentclass[aps,prl,multicol,reprint,preprintnumbers, footbib]{revtex4-2}

\usepackage{amsmath}
\usepackage{graphicx,xcolor}
\usepackage{comment}
\usepackage{amsmath,amssymb,mathtools,slashed}
\usepackage{braket}
\usepackage{feynmp-auto}
\usepackage{simpler-wick}
\usepackage{kantlipsum}
\usepackage{float}
\usepackage{multirow}
\usepackage{rotating}

\usepackage[T1]{fontenc} % if needed
\usepackage{etoolbox}
\usepackage{extarrows}
\usepackage{graphicx}
\usepackage{dcolumn}
\usepackage{bm}
\usepackage{wasysym}
\usepackage{verbatim}

\usepackage{empheq}
\usepackage{multirow}
 
\usepackage{balance}

\usepackage{subfigure}
\usepackage[colorlinks=true,urlcolor=blue,linkcolor=blue]{hyperref}
\usepackage[capitalise]{cleveref}
\usepackage{siunitx}
\usepackage{enumerate}
\newcommand{\be}{\begin{equation}}
\newcommand{\ee}{\end{equation}}

\newcommand{\bdf}{\boldsymbol{f}}

\newcommand{\bn}
{\boldsymbol{n}}
\newcommand{\bdm}
{\boldsymbol{m}}

\begin{document}

\preprint{DESY-26-051}

\title{Squashed Pyramid Interferometer Network (SPIN):\\ Direct Access to Chirality of Cosmological Gravitational Waves}

%\title{Beyond Planar Detectors: Probing Cosmological Chiral Gravitational Waves}

%\title{Detecting Circular Polarization of a Stochastic gravitational wave Background with the Einstein Telescope and L-Shaped Interferometers}

\author{Dmitri E. Kharzeev}
\affiliation{Center for Nuclear Theory, Department of Physics and Astronomy,
Stony Brook University, New York 11794-3800, USA}
\affiliation{Energy and Photon Sciences Directorate, Condensed Matter and Materials Sciences Division, Brookhaven National Laboratory, Upton, New York 11973-5000, USA}

\author{Azadeh Maleknejad}
\affiliation{Centre for Quantum Fields and Gravity, Swansea University,
Swansea SA2 8PP, United Kingdom}
\affiliation{Deutsches Elektronen-Synchrotron DESY,
Notkestra\ss e 85, 22607 Hamburg, Germany}
\affiliation{Institute of Theoretical Physics, Universit\"at Hamburg,
22761 Hamburg, Germany}

\author{Saba Shalamberidze}
\affiliation{Center for Nuclear Theory, Department of Physics and Astronomy,
Stony Brook University, New York 11794-3800, USA}
\affiliation{Physics Department, Brookhaven National Laboratory, Upton, New York 11973-5000, USA}
\date{\today}

%======================================================================
\begin{abstract}

The cosmological gravitational wave background provides a powerful window on parity-violating physics at energies far beyond the reach of terrestrial experiments. However, any colocated planar detector network is insensitive to isotropic circular polarization, independent of its relative orientation. In this Letter, we show that this no-go result can be evaded by a new class of colocated 3D interferometer designs, which we call Squashed Pyramid, whose non-coplanar configuration geometrically isolates chirality. The design can be viewed as a minimal extension of the Einstein Telescope geometry, obtained by introducing a slightly tilted arm relative to the ET planar configuration.  The coplanar correlation channel is blind to circular polarization, whereas the colocated non-coplanar channel is insensitive to the unpolarized background and acquires a response only in the presence of nonzero net helicity. Squashed Pyramid interferometer networks therefore furnish a unique probe of cosmological gravitational wave chirality, opening a realistic terrestrial pathway to test parity violation and fundamental symmetry breaking in the early Universe.

\end{abstract}

\maketitle

%\tableofcontents
%======================================================================
%\section{Introduction}
%\label{sec:intro}

Gravitational waves carry information not only about their astrophysical and cosmological origins, but also about the fundamental interactions of nature, because gravity universally couples to all forms of energy and matter. Their isotropic circular polarization, in particular, would be a uniquely cosmological signature of parity violation in the early Universe, free from astrophysical contamination. However, colocated planar interferometers, such as the Einstein Telescope (ET) triangle, are individually blind to the Stokes-$V$ parameter; even in a network, they are sensitive only to a mixture of the intensity and part of the chiral signal~\cite{Seto:2008sr,Smith:2016jqs,Branchesi:2023,Duval:2025vfg}.

\begin{figure}[t!]    \includegraphics[height=3.7cm]{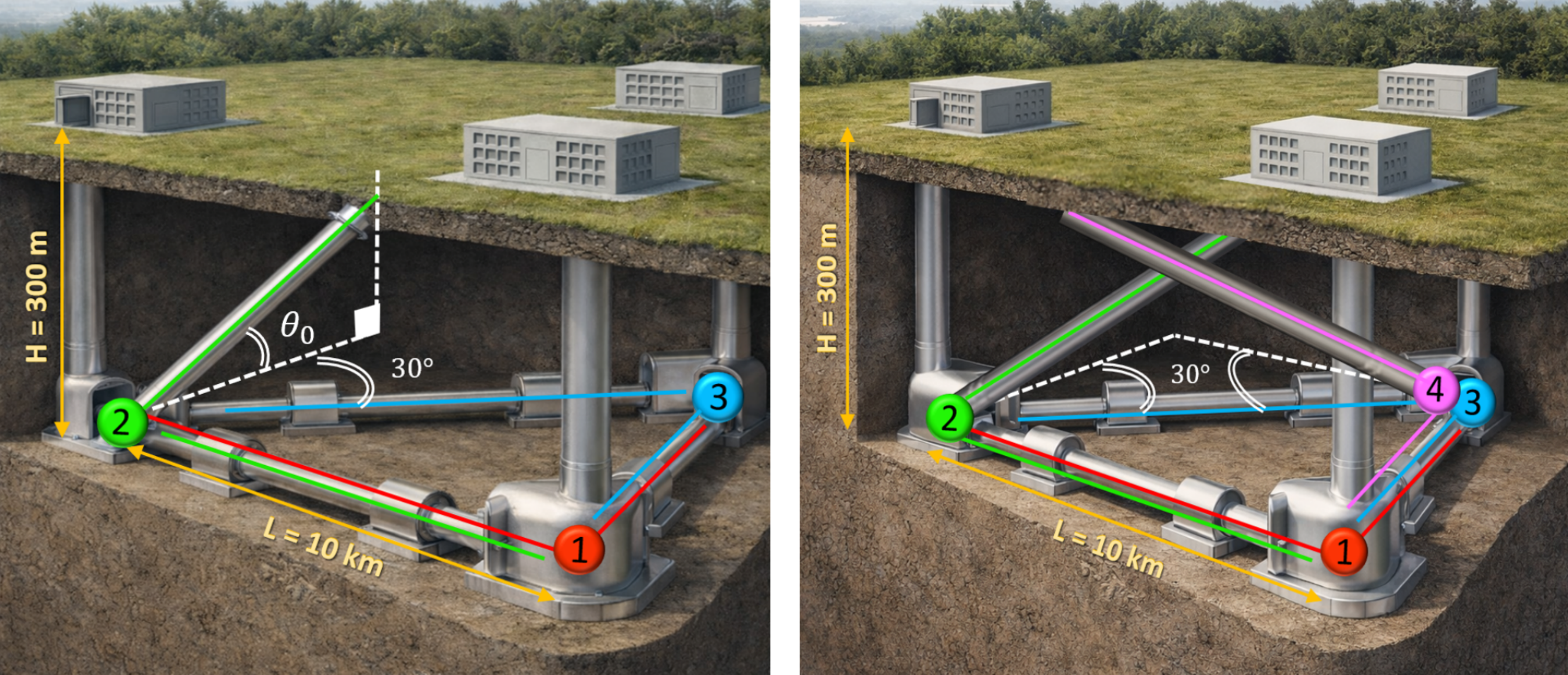} % Right plot
    \caption{Schematic illustration of a 3D interferometer geometry obtained by extending the standard underground triangular Einstein Telescope (ET) layout into a non-coplanar configuration. We consider the ET with arm length $L=10\,\mathrm{km}$ and underground depth $H=300\,\mathrm{m}$. The ET lies in the plane 1-2-3, whereas its non-coplanar extensions are illustrated in the left (\emph{Minimal Pyramid}) and right (\emph{Double Pyramid}) panels. The detectors are shown in red, green, blue, and pink; colored disks denote beam splitters, and lines of the same color denote the corresponding arms. The tilted arm remains underground, with a very mild tilt $\theta_0$ comparable to that of the LHC tunnel at CERN. ($\theta_0 \approx H/L=1.72^\circ$)}
    \label{fig:pyramid}
\end{figure}

In this Letter, we introduce a new class of terrestrial laser-interferometer designs, which we call Squashed Pyramid Interferometer Network (SPIN) (see \cref{fig:pyramid}), whose non-coplanar structure evades this no-go theorem by geometrically isolating chirality. The coplanar channel is blind to circular polarization, while the colocated non-coplanar channel is blind to the unpolarized background and responds only to nonzero net helicity (see \cref{{tab:channels}}).  We refer to this new class of designs as Pyramid configurations, since connecting all the vertices yields a squashed pyramid-like geometry.

Parity violation can arise in a wide range of early-Universe scenarios, reflecting diverse mechanisms through which fundamental symmetries may be broken at high energies. Within this broad landscape, several particularly well-motivated and extensively studied examples have been shown to produce a chiral, or circularly polarized, gravitational wave background. In the inflationary context, such sources arise, for instance, in scenarios where parity violation is generated by Abelian \cite{Cook:2011hg,Anber:2012du} or non-Abelian \cite{Maleknejad:2011jw,Dimastrogiovanni:2012ew,Adshead:2013qp,Maleknejad:2016qjz} gauge-field interactions, or through gravitational interactions in Chern–Simons gravity \cite{Alexander:2004us,Satoh:2007gn}. Post-inflationary processes offer additional avenues for generating chirality. Notable examples include the chiral magnetic effect \cite{Kharzeev:2007jp,Fukushima:2008xe,Kharzeev:2013ffa}, which can trigger plasma instabilities even within the standard cosmological framework \cite{Brandenburg:2023imm}; turbulence in the primordial plasma generated during first-order cosmological phase transitions, such as electroweak or QCD transitions beyond the Standard Model \cite{Witten:1984rs,Kamionkowski:1993fg}; gravitational wave production sourced by primordial magnetic fields coupled to the cosmological plasma \cite{Brandenburg:2021aln}; gauge preheating \cite{Adshead:2015pva}; and audible axions \cite{Machado:2018nqk}. For detailed discussions of parity-violating sources in axion inflation, see \cite{Maleknejad:2012fw,Komatsu:2022nvu}. For a broader overview of other early-Universe sources and mechanisms, see \cite{LISACosmologyWorkingGroup:2022jok,Maleknejad:2025clz}. Parity violation in the Universe can also be probed with the CMB, e.g. cosmic birefringence measurements of its polarization anisotropies \cite{Minami:2020odp}.

\vspace{-0.1cm}
\begin{table}[t!]
\centering
\small
\begin{tabular}{lcc}
\hline
\begin{tabular}{@{}c@{}}Detection\\ pair \end{tabular} & \begin{tabular}{@{}c@{}}Sensitivity\\ $\{\gamma_{_{I}},\gamma_{_{V}}\}$\end{tabular} & \begin{tabular}{@{}c@{}}Channel\\ separation\end{tabular} \\ \hline
ET-ET   &  \quad \rule{0pt}{0.35cm}$\{\frac38,\,0\}$ & \quad pure $I$ \\
ET-PyET & \quad \rule{0pt}{0.35cm} $\{0,\,\frac{\pi}{2} \, \frac{Hf}{c}\}$ & \quad pure $V$ \\
ET-ET 2L   & \quad  $\{\gamma_{_{I}}(f),\,\gamma_{_{V}}(f)\}$ & \quad mixed $I$-$V$ \\
\hline
\end{tabular}
\caption{Network response of detector pairs to the intensity ($I$) and circular-polarization ($V$) components of the SGWB. ET-ET denotes the coplanar triangular Einstein Telescope configuration, PyET the non-coplanar arm, and ET-ET 2L the spatially separated two L Einstein Telescope configuration. }
\label{tab:channels}
\end{table}

\begin{figure}
\includegraphics[height=3cm]{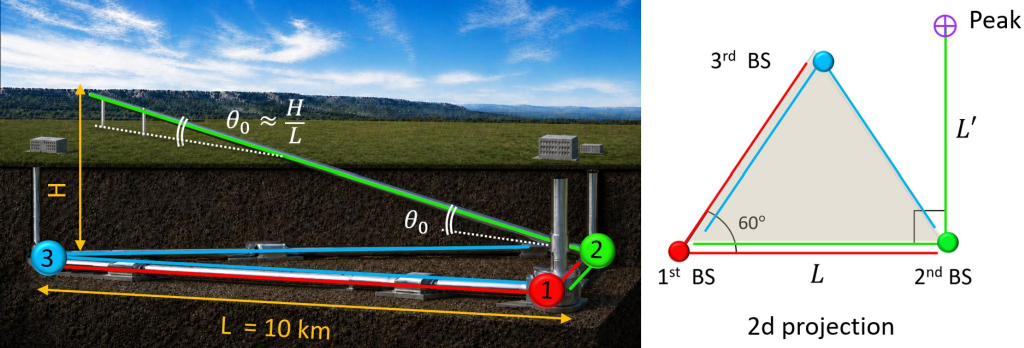}
\caption{Cliff-Extended Squashed Pyramid. This detector geometry extends the minimal 
non-coplanar configuration by continuing the tilted arm toward a natural elevated 
endpoint, such as a mountain slope or cliff face, inspired by landscapes like the Jura cliffs near the France–Switzerland border. The notation is the same as in \cref{fig:pyramid}. 
In the frequency range relevant for our analysis, the sensitivity is dominated by 
quantum shot noise~\cite{Danilishin:2019dxq,LartauxCapocasa:2025,SM}; therefore, 
placing the non-coplanar arm along an elevated mountain or cliff trajectory does 
not lead to a significant loss of sensitivity. The figures are schematic only, 
and the relative dimensions are not to scale. 
 As a representative benchmark, we take $H=1.3\,\mathrm{km}$, corresponding to $\theta_0 \simeq 7.47^\circ$. (BS = beam splitter, $L'/L=\bigl[1-(H/L)^2\bigr]^{1/2}$.)
}
\label{fig:3d-co-}
\end{figure}

The observational significance of chiral gravitational waves lies in their direct sensitivity to fundamental symmetry breaking in the early Universe. Unlike the overall amplitude or spectral shape of a stochastic gravitational wave background, chirality provides a qualitatively distinct observable that cannot be mimicked by parity-symmetric physics or standard astrophysical sources. As such, the detection of circular polarization in gravitational waves would constitute a clear and robust signature of new physics, offering a unique probe of high-energy interactions, cosmic initial conditions, and the symmetry structure of the primordial Universe.

Throughout this Letter, \(H\) denotes the pyramid height, whereas \(H_0\) denotes the Hubble parameter today.

\vskip 0.1cm

\textbf{Framework.}  To set the notation, we briefly review the description of stochastic gravitational wave backgrounds in terms of their two propagating tensor polarizations and summarize how circularly polarized GWs imprint observable signatures in laser interferometers. We consider the perturbed metric of spacetime as
\begin{equation}
ds^2 = -c^2 dt^2 + \bigl[\delta_{ij} + h_{ij}(t,\mathbf{x})\bigr] dx^i dx^j,
\label{eq:metric}
\end{equation}
where $h_{ij}(t,\mathbf{x})$ denotes the GW perturbation,
treated at linear order. A stochastic gravitational wave background may be decomposed into circular
polarization modes as
\begin{equation}
h_{ij}(t,\mathbf{x})
=
\sum_{\sigma=R,L}\int_{-\infty}^{+\infty} df\; \int d\Omega_{\bdf} \, 
\tilde{h}_\lambda(\bdf)\, e^{\lambda}_{ij}(\hat{\bdf})
e^{-i\varphi},
\label{eq:hij_circular}
\end{equation}
where $\hat \bdf$ denotes the propagation direction of GWs and $\varphi \equiv 2\pi f (t - \frac{\hat{\bdf}\cdot\mathbf{x}}{c})$. The statistical properties of a generic isotropic SGWB are encoded in the coherency matrix
\begin{equation}
\big\langle
\tilde{h}_A(\bdf)\, \tilde{h}_B^\ast(\bdf')
\big\rangle
=
\frac{1}{2}\,
\delta(f-f')\,
\delta^2(\hat{\bdf}-\hat{\bdf}')\;
\mathcal{S}_{AB}(f),
\label{eq:power-GW}
\end{equation}
where $A,B\in\{R,L\}$ and $\mathcal{S}_{AB}(f)$ is the spectral density matrix. The polarization content of the SGWB is conveniently characterized by the
Stokes parameters,
\begin{align}
I(f)
&\equiv
\mathcal{S}_{RR}(f)
+
\mathcal{S}_{LL}(f),\\
V(f)
&\equiv
\mathcal{S}_{LL}(f)
-
\mathcal{S}_{RR}(f),
\end{align}
where $I$ measures the total SGWB intensity, whereas $V$ characterizes its net circular polarization. Being parity odd, $V$ probes chiral GW sources and parity-violating dynamics. In cosmology, the SGWB is conveniently characterized by the dimensionless energy density per logarithmic frequency interval, $\Omega_{\rm GW}(f)$, related to the Stokes intensity $I(f)$ by
\begin{align}
\Omega_{\rm GW}(f)=\frac{c^2\pi}{2G \rho_c}\, f^3 \, I(f),
\end{align}
where $\rho_c=3H_0^2/(8\pi G)$ is the critical density of the universe today. We also define the polarization degree
\begin{align}
\Pi(f)=\frac{V(f)}{I(f)}.
\end{align}

For a Michelson laser interferometer with arm directions $\hat{\bn}$ and $\hat{\bdm}$, the differential phase delay measured by a detector at $\boldsymbol{x}_A$ is given by
\begin{align}\nonumber
\Delta\Phi(t)=\frac{2\omega_0 L}{c}
\sum_{\lambda}
\int_{-\infty}^{\infty} &df \, \int d\Omega_{\bdf} \,\tilde{h}_\lambda(\boldsymbol{f}) \\&
\times\,
\mathcal{F}^\lambda_{\bn,\bdm}(\hat{\boldsymbol{f}})
\,e^{\,2i\pi(\frac{\boldsymbol{f}\cdot\boldsymbol{x}_A}{c}-f t)},
\end{align}
where the circular antenna pattern
$\mathcal{F}^{\lambda}_{\bn,\bdm}$ can be decomposed as $\mathcal{F}^{\lambda}_{\bn,\bdm}=A_{\bn,\bdm}-i \sigma_\lambda B_{\bn,\bdm}$ with
\begin{align}
    A_{\bn,\bdm}&=\frac{1}{2\sqrt2}\left[(n_\theta^2-n_\phi^2)-(m_\theta^2-m_\phi^2)\right] ,\\
    B_{\bn,\bdm}&=\frac{1}{\sqrt2}\left(n_\theta n_\phi-m_\theta m_\phi\right).
\end{align}
The triangular Einstein Telescope provides three colocated, differently oriented interferometers, enabling internal cross-correlations that enhance sensitivity to stochastic backgrounds and suppress uncorrelated noise \cite{Branchesi:2023}. For two interferometers with arm directions $(\hat{\bn},\hat{\bdm})$ and $(\hat{\bn}',\hat{\bdm}')$, separated by $\Delta\mathbf{X}$, the phase correlation for an isotropic SGWB is
\begin{align}
\langle\Delta\Phi_1 \Delta\Phi_2 \rangle  = \frac{8\omega_0^2L^2}{5\pi c^2} 
\int df 
\big( \gamma_{_{I}} I_0(f)  + \gamma_{_{V}}  V_0(f) \big),
\end{align}
where $\gamma_I$ and $\gamma_V$ are the overlap reduction functions for the intensity and circular polarization, respectively, given in the $\frac{f|\Delta\mathbf{X}|}{c}<<1$ regime by 
\begin{align}
\gamma_{_{I}}
    &= \frac{5}{4\pi} \int d\Omega_{\hat\bdf} \,
    \bigl( A A' + B B' \bigr),
    \label{eq:Gamma--i}\\
    \gamma_{_{V}}
    &= \frac{5}{2c}  \int d\Omega_{\hat\bdf} \;
    \bdf\cdot\!\Delta \boldsymbol{X} \,
    \bigl( A B' -  A' B \bigr).
    \label{eq:Gamma--v}
\end{align}
A necessary condition for sensitivity to the circular polarization (Stokes-$V$) component of an isotropic SGWB is that the detector network be genuinely three dimensional. In particular, $\gamma_V$ vanishes for either $\Delta \mathbf{X}=0$ or a coplanar configuration.

\vskip 0.1cm
\textbf{Squashed Pyramid interferometers and sensitivity to Stokes-$V$.}  We introduce a new class of detector configurations, dubbed Squashed Pyramid interferometers, in which non-coplanar interferometers remain colocated within a genuinely three-dimensional network. This geometric extension lifts the planar suppression of the circular response and enables sensitivity to the Stokes-$V$ component of an isotropic SGWB. As we show, Squashed  Pyramid interferometers admit a nonvanishing parity-odd response, thereby providing a minimal framework for probing chiral SGWB and the full three-dimensional tensor structure of the GW field.

Here we consider two representative realizations of this class of non-coplanar, colocated designs, naturally motivated by next-generation underground gravitational wave observatories. The first is a \emph{Minimal Squashed  Pyramid} configuration: a non-planar extension of the Einstein Telescope, obtained by adding a single tilted arm, displaced by a height $H$ from the detector plane, to the triangular ET layout, as shown in the left panel of \cref{fig:pyramid}. The second is a \emph{Double Squashed  Pyramid} configuration, in which two arms remain in the ET plane and are supplemented by two additional tilted arms, as shown in the right panel of \cref{fig:pyramid}. Finally, this tilt can be further extended by prolonging the tilted arm through a vertical cliff or a     mountain, i.e. \emph{Cliff-Extended Squashed  Pyramid}. We take the cliff height to be 1km, comparable to the Jura cliffs near the France–Switzerland border, not as a near-term design target, but rather as a concrete benchmark, as illustrated in \cref{fig:3d-co-}. Throughout this Letter, we use ET to denote the coplanar interferometer arms of the Einstein Telescope, while PyET refers to the non-coplanar arm in the Squashed Pyramid Einstein Telescope configuration.

Let us focus on the minimal Pyramid configuration, characterized by the interferometer arm length $L$ and the Pyramid height $H$. The detectors are specified as
\begin{equation}
\begin{array}{lll}
 \boldsymbol{D}: ~~~ \{\hat{\bn}= (1,0,0), \hat{\bdm}=(\frac12,\frac{\sqrt3}{2},0)\},
 \\  \boldsymbol{D}': ~ \{\hat{\bn}'=(\frac12,-\frac{\sqrt3}{2},0), \hat{\bdm}'=(-\frac12,-\frac{\sqrt3}{2},0)\},
\\  \boldsymbol{D}'': ~~ \{\hat{\bn}''=\lambda (-\frac{\sqrt3}{2},-\frac12,\frac{H}{L}),
 \hat{\bdm}''=(-\frac12,\frac{\sqrt3}{2},0)\},
\end{array}
\end{equation}
where $\lambda=(1+(\frac{H}{L})^2)^{-1/2}$. 
The displacement vectors connecting the beam splitters are
$\Delta \boldsymbol{X}_{12} = -L\,\hat{\bdm}$, $\Delta \boldsymbol{X}_{13} = -L\,\hat{\bn}$, and $\Delta \boldsymbol{X}_{23} = -L\,\hat{\bdm}''$.
For the coplanar pair $(D,D')$, corresponding to the standard (ET-ET) configuration, the overlap reduction functions are
\begin{equation}
\gamma_{_{I}}(f) = \frac{3}{8} , \qquad \gamma_{_{V}}(f) = 0 , \qquad \text{(ET-ET)}
\end{equation}
so this channel is sensitive to the intensity $I$ but blind to the circular polarization $V$. By contrast, for the non-coplanar pair $(D,D'')$, corresponding to the (ET-PyET) network, we find
\begin{equation}
\tilde{\gamma}_{_{I}}(f) = 0 , \qquad \tilde{\gamma}_{_{V}}(f) = \frac{\pi}{2} \, \frac{Hf}{c}, \qquad \text{(ET-PyET)}
\label{eq:tilde--}
\end{equation}
so this channel is blind to $I$ and responds only to $V$. %The second coplanar pair $(D',D'')$ has
%\begin{align}
%    \tilde{\gamma}_{_{I}}(f) = \frac{3}{4} , \qquad \tilde{\gamma}_{_{V}}(f) = -\frac{\pi}{2} \, Hf, \qquad \text{(ET-PyET)'}.
%\end{align}
Pyramid interferometers therefore provide a unique probe of chirality in cosmological SGWB, opening a realistic terrestrial avenue for testing parity violation and fundamental symmetry breaking in the early Universe.

To project out the intensity response and isolate the circular-polarization mode at the network level, we define the projected $V$-mode overlap function \cite{SetoTaruya:2007}
\begin{align}
\Gamma_{_{V,\alpha\beta}} \equiv
\frac{\gamma_{_{V,\beta}}\gamma_{_{I,\alpha}}-\gamma_{_{V,\alpha}}\gamma_{_{I,\beta}}}
{(\gamma_{_{I,\alpha}}^2+\gamma_{_{I,\beta}}^2)^{1/2}} .
\end{align}
It is constructed so that the intensity contribution cancels, leaving a response only to the $V$ mode. In the ideal case $\gamma_{_{V,\alpha}}=0$ and $\gamma_{_{I,\beta}}=0$, one finds $\Gamma_{_{V,\alpha\beta}}=\pm\gamma_{_{V,\beta}}$,  corresponding to maximal separability of the stochastic $I$ and $V$ channels: one baseline probes only intensity, while the other responds only to circular polarization. This shows that the PyET provides an optimal design for isolating the circular-polarization signal of the SGWB. A possible non-colocated ET--CE network, with detectors placed at specific terrestrial sites and requiring an additional detector to achieve $\gamma_{_{I}}=0$, was discussed in \cite{SetoTaruya:2007}.

\vskip 0.1cm

\textbf{Sensitivity and dominant noise sources.} Various noise sources limit the sensitivity of interferometric GW detectors \cite{Danilishin:2019dxq, Branchesi:2023}. A key contribution is quantum noise, which originates from the quantum nature of light and its interaction with the interferometer mirrors, and manifests in two complementary forms. At high frequencies, the sensitivity is limited by shot noise (phase noise), arising from the quantum uncertainty in the number of photons detected over a finite time interval. At low frequencies, quantum radiation pressure noise (amplitude noise) becomes dominant, reflecting fluctuations in the radiation pressure exerted by photons on the test masses. The interplay between these two contributions sets the fundamental quantum limit on detector sensitivity across a large fraction of the observational band.

The Einstein Telescope adopts a xylophone configuration, in which each interferometric site hosts two complementary detectors optimized for different frequency ranges \cite{Branchesi:2023}. The low-frequency instrument (LF-ET) is designed to operate in the $1-30\mathrm{Hz}$ band, employing cryogenic mirrors, low optical power, and advanced seismic isolation to suppress thermal and environmental noise. In contrast, the high-frequency instrument (HF-ET) targets the $30\mathrm{Hz}$–10 kHz regime, utilizing high laser power and room-temperature optics to reduce quantum shot noise. This separation mitigates otherwise unavoidable trade-offs between thermal and quantum noise, enabling broadband sensitivity that cannot be achieved with a single interferometer design. Current interferometric technology remains effective up to frequencies of order $10^4$ Hz. However, several new laboratory-scale concepts have been proposed to extend the GW window to higher frequencies \cite{Aggarwal:2025noe}, including quantum-enhanced strategies~\cite{Kharzeev:2025lyu}.

A possible concern in the tilted-arm configuration is that the tilt introduces a vertical-to-longitudinal projection of suspension motion proportional to  \(H/L\). However, this effect enters the cross-correlation only through the small residual vertical coupling of the planar ET arms. Consequently, the corresponding vertical-noise two-point function remains small; see the Supplemental Material for details on vertical suspension noise in the tilted-arm configuration~\cite{SM}. 

Another possible concern about the tilted geometry is its impact on seismic and Newtonian, or gravity-gradient, noise. In fact, a primary motivation for placing the Einstein Telescope underground is precisely the strong suppression of these noise sources, which dominate the low-frequency end of the observational band. At higher frequencies, however, \(f\sim 30\!-\!10^{4}\,\mathrm{Hz}\), the sensitivity is instead limited mainly by quantum shot noise. Therefore, introducing non-coplanar PyET interferometer arms at different vertical levels is not expected to incur a significant noise penalty, while enabling access to additional polarization observables. See the Supplemental Material \cite{SM} for a discussion of long-distance seismic correlations.

\begin{figure}
    \centering
    \includegraphics[width=1\linewidth]{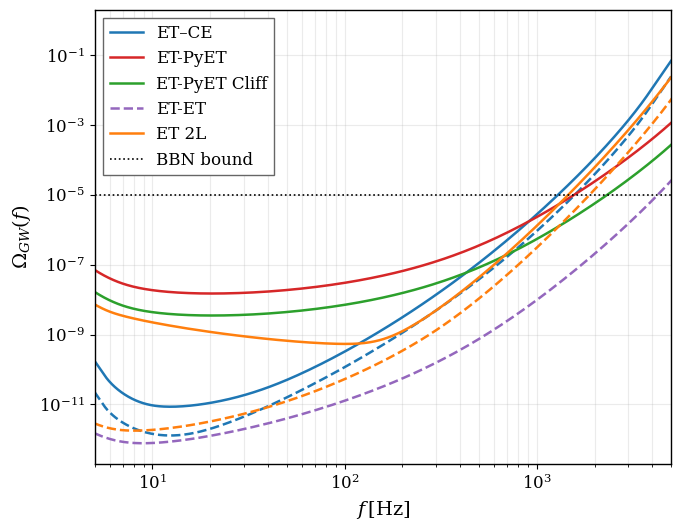}
    \caption{Power-law sensitivity curves for stochastic gravitational wave backgrounds for three years of observation and SNR=1. We compare ET–CE and ET–ET detector pairs, the 2L ET configuration, and the non-coplanar Pyramid configurations in their minimal (ET–PyET) and cliff-extended (ET–PyET cliff) realizations. The ET site is taken to be Euregio Meuse--Rhine (EMR), with a 20-km CE detector at Livingston; for the 2L configuration, the second ET is placed in Sardinia with its orientation rotated by 45$^\circ$ relative to the EMR site, as in~\cite{Branchesi:2023}. Dashed lines denote sensitivity to the Stokes-$I$ component and solid lines to the Stokes-$V$ component. The big-bang nucleosynthesis (BBN) bound is shown as a dotted line.}
    \label{fig:Omega-GW}
\end{figure}

Here we compute the signal-to-noise ratio (SNR) for pairs of interferometric detectors. The signal $s_\alpha=\frac{c}{2\omega_0L}\Delta\Phi_\alpha$ in detector $\alpha$ is $s_\alpha(t)
=
h_\alpha(t)
+
n_\alpha(t)$, where $h_\alpha$ and $n_\alpha$ are the gravitational wave signal and detector noise respectively. The noise spectrum is 
\begin{equation}
\big\langle
 n_\alpha(f)n^\ast_{\beta}(f')
\big\rangle
=
\frac{\delta_{\alpha\beta}}{2}\,
\delta(f-f')
\mathcal{N}_{\alpha}(f),
\end{equation}
where $\mathcal N_\alpha(f)$ is the strain-equivalent noise spectral density of detector $\alpha$.
\color{black}

The SNR can be computed through
\begin{align}
\mathrm{SNR}
&=
\left(\frac{3H_0^2}{10\pi^2}\right)
\left(
2T_{\rm obs}
\int_0^{\infty} df\,
\frac{\Omega_{\rm GW}^2(f)}{f^6}
\right.
\nonumber\\
&\qquad\qquad\left.
\times
\frac{\big(\gamma_{_{I}}(f)+\gamma_{_{V}}(f)\,\Pi(f)\big)^2}
{\mathcal{N}_{\alpha}(f)\,\mathcal{N}_{\beta}(f)}
\right)^{\frac12},
\end{align}
where $T_{\rm obs}$ is the observation time. We use this expression to evaluate the sensitivity of the Pyramid configuration to the Stokes-V component. To place these results in context, we compare with correlations involving the Cosmic Explorer (CE), a proposed U.S. third-generation ground-based gravitational wave observatory~\cite{Evans:2021gyd}. We further compare with the sensitivity for the 2L configuration of ET~\cite{Branchesi:2023}, an alternative to the triangular ET design in which the single triangular detector is replaced by two L-shaped interferometers at different sites. 

A useful tool in the study of the sensitivity of a network of detectors to isotropic GWB is the power-law sensitivity (PLS) curve \cite{Thrane:2013oya}.
Using the above SNR expressions, we present the corresponding PLS sensitivities for the ET-ET, ET-PyET, ET 2L configuration, and ET-CE networks in \cref{fig:Omega-GW}; the main results are summarized in Table~\ref{tab:channels}. Relative to the minimal Squashed  Pyramid, the double-pyramid configuration provides two independent parity-sensitive correlation channels. Their contributions add in quadrature, yielding an overall improvement of the circular-polarization signal-to-noise ratio by a factor of $\sqrt{2}$.

\vskip 0.1cm 
\textbf{Discussion and outlook.} Detecting a stochastic gravitational wave background would open a unique window on the early Universe, while its circular polarization would directly probe primordial parity violation and chiral dynamics. Yet colocated planar interferometers are intrinsically blind to the isotropic circularly polarized component, blocking access to this key observable. Although a non-colocated pair of third-generation detectors, such as Einstein Telescope and Cosmic Explorer, can become sensitive to Stokes \(V\), their response generically mixes the intensity and chiral components, preventing a clean separation of the stochastic \(I\) and \(V\) channels. In addition, this strategy relies on the construction of a separate laser-interferometric detector with third-generation sensitivity, comparable to ET, which is a major experimental investment. In this Letter, we introduce a new class of detector designs, which we call \emph{Squashed Pyramid Interferometer Network} (SPIN), that evade this no-go result (see \cref{tab:channels}). Our proposal is substantially more economical, requiring only the addition of a tilted arm to an ET-like infrastructure rather than a distinct new observatory (see \cref{fig:pyramid,fig:3d-co-}).  

We show that Squashed Pyramid interferometers geometrically isolate chirality: the coplanar correlation channel is blind to circular polarization, while the colocated non-coplanar channel is blind to the unpolarized background and acquires a response only in the presence of nonzero net helicity (see \cref{eq:tilde--}). That identifies the Pyramid configuration as the optimal design for isolating the circular-polarization signal of the SGWB.

In this Letter, we introduced a class of non-coplanar, colocated interferometers naturally compatible with next-generation underground observatories, and presented three representative realizations of this concept: i) a Minimal Squashed Pyramid (left panel of \cref{fig:pyramid}), obtained by adding a single tilted arm to the triangular Einstein Telescope; ii) a Double Pyramid (right panel of \cref{fig:pyramid}), in which the ET plane is supplemented by two tilted arms; and iii) a futuristic Cliff-Extended Minimal Pyramid (\cref{fig:3d-co-}), where the tilted arm is further prolonged through a vertical cliff or mountain. In the minimal case the tilted arm remains underground, with a very mild tilt comparable to that of the LHC tunnel at CERN. The tilted arm does not introduce a significant noise penalty: vertical suspension effects are suppressed, and underground seismic/Newtonian correlations remain small; see SM \cite{SM}. Compared to a 2L ET configuration, the Squashed Pyramid preserves the redundancy and null-stream structure of the triangular ET geometry, while its single tilted arm breaks coplanarity and provides direct access to the isotropic Stokes-$V$ component. The corresponding $I$- and $V$-sensitivities of the $10\,\mathrm{km}$ triangular Squashed Pyramid and the $15\,\mathrm{km}$ 2L ET configuration are compared in \cref{fig:Omega-GW}.

Pyramid interferometers open a new observational window onto parity violation and chiral symmetry breaking in the fundamental interactions, as imprinted on gravitational waves. Their realization would establish a realistic terrestrial probe of early-Universe physics, with sensitivity to inflationary dynamics, axion--gauge-field sectors, modified gravity, and a broad class of primordial chiral sources, including the chiral magnetic effect, primordial turbulence, and primordial magnetic fields. More broadly, our results identify colocated non-coplanarity as a fundamental geometric principle for the design of future interferometric detectors targeting cosmological chirality.

\makeatletter
\let\latex@addcontentsline\addcontentsline
\renewcommand{\addcontentsline}[3]{}
\makeatother

\vskip 0.2cm 
\begin{acknowledgments}
{\bf{Acknowledgments:} } We thank Oliver Gerberding, Eiichiro Komatsu, and Geraldine Servant for their valuable comments on the draft. We are grateful to Simone Blasi, Chiara Caprini, Valerie Domcke, John Ellis, and Thomas Konstandin for helpful discussions. AM gratefully acknowledges the warm hospitality of DESY and Geraldine Servant during the completion of this work. The work of AM was supported by the Royal Society through a University Research Fellowship (Grant No. RE22432) and by the Deutsche Forschungsgemeinschaft under Germany’s Excellence Strategy (EXC 2121 Quantum Universe – 390833306). The work of D.K. and S.S. was supported by the U.S. Department of Energy, Office of Science, Office of Nuclear Physics, Grants No. DE-FG88ER41450 and DE-SC0012704 and by the U.S. Department of Energy, Office of Science, National Quantum Information Science Research Centers, Co-design Center for Quantum Advantage (C2QA) under Contract No.~DE-SC0012704. The work of S.S. was also supported by the U.S. Department of Energy, Office of Science, Office of Workforce Development for Teachers and Scientists, Office of Science Graduate Student Research (SCGSR) program. The SCGSR program is administered by the Oak Ridge Institute for Science and Education (ORISE) for the DOE. ORISE is managed by ORAU under contract number DESC0014664. All opinions expressed in this paper are the authors' and do not necessarily reflect the policies and views of DOE, ORAU, or ORISE.
\end{acknowledgments}

\makeatletter
\let\addcontentsline\latex@addcontentsline
\makeatother

%================ Supplemental Material ====================

\clearpage
\onecolumngrid

%-----------------------------------------------------------
% Reset counters and switch to S-numbering
%-----------------------------------------------------------

\setcounter{page}{1}
\renewcommand{\thepage}{S\arabic{page}}

\setcounter{section}{0}
\renewcommand{\thesection}{S\arabic{section}}

\setcounter{equation}{0}
\renewcommand{\theequation}{S\arabic{equation}}

\setcounter{figure}{0}
\renewcommand{\thefigure}{S\arabic{figure}}

\setcounter{table}{0}
\renewcommand{\thetable}{S\arabic{table}}

%-----------------------------------------------------------
% Hyperref anchors (avoid duplicate warnings)
%-----------------------------------------------------------

\makeatletter
\renewcommand{\theHsection}{S\arabic{section}}
\renewcommand{\theHequation}{S\arabic{equation}}
\renewcommand{\theHfigure}{S\arabic{figure}}
\renewcommand{\theHtable}{S\arabic{table}}
\makeatother

%===========================================================
% Supplemental title
%===========================================================

\begin{center}
\textbf{\large Supplemental Material for:\\[4pt]
\emph{Squashed Pyramid Interferometer Network (SPIN):\\ Direct Access to Chirality of Cosmological Gravitational Waves}}\\[8pt]
Dmitri E. Kharzeev, Azadeh Maleknejad, and Saba Shalamberidze
\end{center}

\vspace{8pt}

%===========================================================
% Supplemental abstract/introduction
%===========================================================

For completeness, in Sec.~S.1 we briefly review the basic design of the Einstein Telescope, the relevant detector noise budget, and the response of interferometer networks to a stochastic gravitational-wave background (SGWB). In Sec.~S.2, we focus on the tilted-arm configuration and estimate the associated vertical suspension noise relevant for SGWB searches. Finally, in Sec.~S.3, we discuss long-distance seismic correlations and explain why the tilted arm does not significantly affect the SGWB sensitivity.

\vspace{10pt}

%===========================================================
% Remove unwanted entries from TOC
%===========================================================

\addtocontents{toc}{\protect\setcounter{tocdepth}{1}}

%===========================================================
% Table of contents
%===========================================================

\tableofcontents

\vspace{12pt}

%===========================================================
% Supplemental sections
%===========================================================
\section{S1. The Einstein Telescope and Its Noise Budget}

\begin{figure*}[h]
\centering
\includegraphics[width=0.95\linewidth]{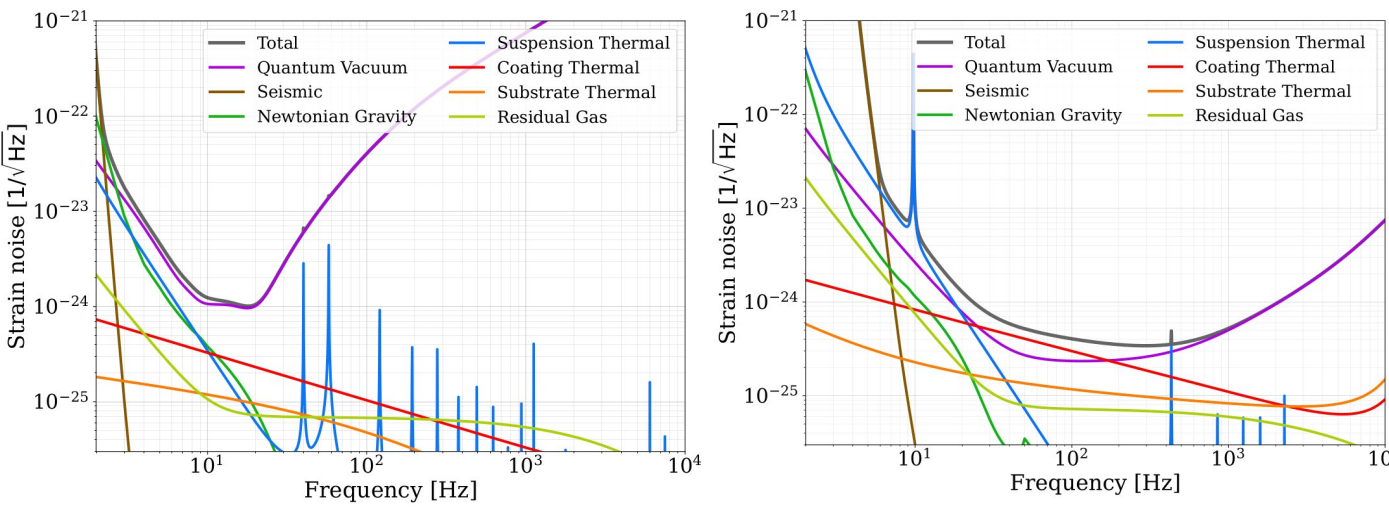}
\caption{Noise budget of the Einstein Telescope. The figures show the contributions of the main known noise sources to the detector strain sensitivity in the low-frequency configuration (ET-LF) (left panel) and high-frequency configuration (ET-HF) (right panel), including quantum noise, seismic and Newtonian noise, thermal noises, and residual gas. Plot credit: \cite{LartauxCapocasa:2025}.}
 \label{fig:noise}
\end{figure*}

 Interferometric GW detectors are affected by a variety of noise sources, as illustrated by the known Einstein Telescope noise budget in \cref{fig:noise}. At low frequencies, seismic noise and Newtonian, or gravity-gradient, noise arise from ground motion and fluctuating terrestrial gravitational fields. Thermal noise, associated with the mirrors, coatings, and suspension systems, also contributes significantly across the intermediate band. In the frequency range relevant for our analysis, however, the dominant limitation comes from quantum noise \cite{Danilishin:2019dxq, Branchesi:2023}. These quantum fluctuations are tied to the laser field and its interaction with the interferometer mirrors. At high frequencies, photon-counting uncertainty produces shot noise, appearing as phase fluctuations in the measured signal. At low frequencies, fluctuations in the radiation pressure exerted by photons on the test masses generate radiation-pressure noise. The balance between these two effects sets the fundamental quantum sensitivity limit over much of the observational band.

\begin{figure}[h]
    \centering
    \includegraphics[width=0.3\linewidth]{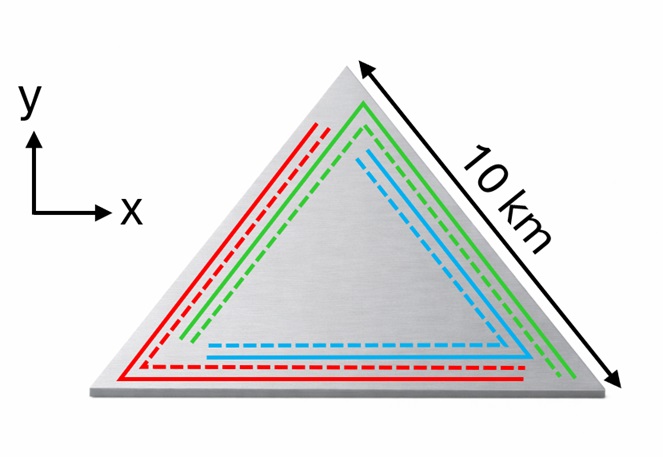}
    \caption{Baseline Einstein Telescope triangular configuration, consisting of three colocated $60^\circ$-opening L-shaped Michelson interferometers. The interferometers are planned to be constructed of order 100–300 m below the Earth’s surface to suppress seismic and environmental noise, as discussed in the ET design and site studies \cite{Branchesi:2023}. The lines represent the laser-beam trajectories of the two colocated interferometers. Solid lines correspond to the high-frequency interferometer operating at $1550\,\mathrm{nm}$, while dashed lines denote the low-frequency interferometer operating at $1064\,\mathrm{nm}$. }
    \label{fig:ET}
\end{figure}

To address these different noise contributions, the Einstein Telescope employs a xylophone configuration, where each site hosts two interferometers optimized for distinct frequency ranges \cite{Branchesi:2023}. The low-frequency detector, ET-LF, is designed for the $1$--$30~\mathrm{Hz}$ range and uses cryogenic mirrors, low circulating optical power, and sophisticated seismic isolation systems to reduce seismic, thermal, and environmental noise. The high-frequency detector, ET-HF, targets the $30~\mathrm{Hz}$--kHz band, using high laser power together with room-temperature optics to improve the shot-noise-limited sensitivity. This division alleviates the otherwise unavoidable tension between thermal and quantum noise requirements, allowing broadband sensitivity that would be difficult to realize within a single interferometer design; see \cref{fig:noise}. One of the main reasons for placing the Einstein Telescope underground is to reduce seismic and Newtonian, or gravity-gradient, noise, which dominate the low-frequency part of the sensitivity curve, as shown in \cref{fig:noise}. At higher frequencies, however, $f\sim 30\!-\!10^{4}\,\mathrm{Hz}$, the limiting contribution is instead quantum shot noise. In this range, low-frequency environmental noise is strongly suppressed and does not set the relevant sensitivity limit. The performance is therefore mainly controlled by quantum noise, implying that adding non-coplanar, tilted interferometer arms at different vertical levels does not introduce a significant noise cost, while opening access to additional polarization observables.

\section{S2. Vertical Suspension Noise in the Tilted-Arm Configuration}\label{S2}

In standard ground-based interferometers such as LIGO and ET, the detector arms lie in a horizontal plane. As a consequence, vertical displacement of the test masses does not project onto the longitudinal arm direction at leading order. Vertical seismic and suspension motion are therefore largely decoupled from the GW readout and are typically neglected, or treated as subleading contributions, in the noise budget. In contrast, in a non-planar configuration where one arm is tilted by a small angle with respect to the horizontal plane, vertical motion acquires a finite projection along the optical axis. Even for small tilts, this leads to a direct coupling of vertical displacement noise into the interferometer output. In such geometries, vertical isolation noise becomes a relevant contribution and must be included in the analysis.

In this section, we compute the contribution of vertical displacement noise associated with the tilted arm to the detector output and its cross-correlation with a standard horizontal detector. We work perturbatively in the small parameter 
\be
\hat n''\cdot \hat z = \sin\theta_0 \approx  H/L \ll 1,
\ee 
where \(H\) characterizes the vertical displacement of the tilted arm and \(L\) is the arm length.
For the end test mass of the tilted arm, we consider a wedged mirror rather than a mechanically tilted mirror. The test mass is suspended in its standard vertical equilibrium configuration, so that gravity keeps the mirror body aligned with the local vertical direction. The optical surface, however, is wedged by a small angle \(\theta_0\). This wedge is chosen such that an incoming beam propagating along the tilted arm is retro-reflected back along the same tilted trajectory. In this way, the geometric tilt of the optical path is implemented optically, through the wedge angle, rather than mechanically, through a tilted suspension or a tilted mirror body. This distinction is important: the suspension remains essentially the standard vertical one, while the beam direction has a small vertical component. Consequently, vertical displacement noise couples to the interferometric phase only through the projection of the mirror motion onto the tilted optical axis.

\begin{figure}[h]
    \centering    \includegraphics[width=0.6\linewidth]{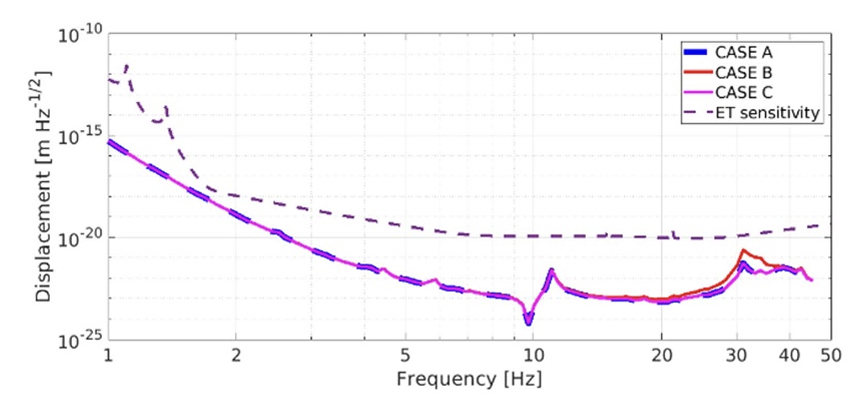}
    \caption{ The expected residual motion of the test mass due to vertical
    ground displacement in the region $1$--$50\,\mathrm{Hz}$. In the region
    $1$--$3\,\mathrm{Hz}$, the test-mass displacement is about one order of
    magnitude below the ET sensitivity curve. In the range
    $5$--$20\,\mathrm{Hz}$, the test-mass displacement is more than two orders
    of magnitude below the ET sensitivity. Around $30\,\mathrm{Hz}$, a mechanical
    resonance of the standard filters reaches the value of
$5\times10^{-19}\,\mathrm{m}/\sqrt{\mathrm{Hz}}$.
    Figure adapted from Ref.~\cite{Bertocco:2024}. }
    \label{fig:vertical-noise-ET}
\end{figure}

For the ideally horizontal ET detectors $D$ and $D'$, the direct projection of
vertical motion onto the optical arms vanishes, i.e.
\be
\epsilon_n=\epsilon_m=0,
\qquad \text{(ideal ET)} .
\ee
Thus, in the ideal horizontal limit,
\(S^{(z)}_{n_D}(f)=S^{(z)}_{n_{D'}}(f)=0\).
In a realistic instrument, due to the Earth's curvature, the coupling coefficient
\(\epsilon\) between the longitudinal and vertical directions is not negligible.
For a \(10\,{\rm km}\) arm, the longitudinal--vertical coupling caused by the
Earth's curvature is \cite{Bertocco:2024}
\be
\epsilon_n(f)=\epsilon_m(f)=\epsilon \simeq 3\times 10^{-3},
\qquad \text{(realistic ET)} ,
\ee
corresponding to an optimistic but realistic level of residual vertical-to-longitudinal
coupling in a well-aligned ET-like detector. The interferometer is sensitive only to the relative vertical motion. For a given
detector \(D\), we define
\be
\Delta z_n(t) \equiv z_E(t)-z_B(t),
\ee
where \(z_E(t)\) and \(z_B(t)\) denote the vertical displacements of the end
mirror \(E\) and the beam splitter \(B\), respectively. In the frequency domain
we describe this motion by its amplitude spectral density, $\left[\sqrt{S_{\Delta z_n}(f)}\right]
= {\rm m}/\sqrt{\rm Hz}$.
In the standard planar ET configuration, this vertical motion enters the
interferometer readout only through the small residual vertical-to-longitudinal
coupling \(\epsilon_n\). The corresponding optical path-length noise amplitude
spectral density is therefore
\be
\sqrt{S^{(z)}_{\delta L_n}(f)}
=
\epsilon_n(f)\,
\sqrt{S_{\Delta z_n}(f)}.
\ee
The resulting contribution of this vertical noise remains well below the ET sensitivity curve
across the relevant frequency range. In \cref{fig:vertical-noise-ET}, we present the expected residual
test-mass displacement due to vertical ground motion for ET ($\sqrt{S^{(z)}(f)}$ for coplanar arms), based on the
NIP-SA suspension study of Ref.~\cite{Bertocco:2024}. The plotted quantity
corresponds to the residual motion after propagation through the vertical
isolation system, in the frequency range $1$--$50\,\mathrm{Hz}$. The three
curves, Cases A--C, represent different NIP-SA configurations with varying
suspension height and total suspended mass. Case A considers an
$8\,\mathrm{m}$ tall suspension with total mass $2650\,\mathrm{kg}$, while
Cases B and C use a $10\,\mathrm{m}$ tall suspension, with Case C increasing
the total mass to $3250\,\mathrm{kg}$. The figure shows that the residual
test-mass displacement lies about one order of magnitude below the ET
sensitivity curve at $1$--$3\,\mathrm{Hz}$ and more than two orders of
magnitude below it in the $5$--$20\,\mathrm{Hz}$ band, with a mechanical
resonance around $30\,\mathrm{Hz}$ reaching
$5\times10^{-19}\,\mathrm{m}/\sqrt{\mathrm{Hz}}$.

In the tilted-arm configuration, the vertical displacement additionally projects geometrically onto the optical axis with a factor $H/L$. Consequently, the induced longitudinal noise is enhanced relative to the standard ET case by approximately $(H/L)/\epsilon$
\be
\sqrt{S^{(z)}_{\delta L_{n''}}(f)}
=
(\epsilon(f)+\frac H L)\,
\sqrt{S_{\Delta z_{n''}}(f)}.
\ee
For the two representative geometries considered here, this gives
\be
\frac{\sqrt{S^{(z)}_{\text{tilt}}(f)}}{\sqrt{S^{(z)}_{\text{ET}}(f)}}
\simeq \frac{H/L}{\epsilon} =
\begin{cases}
10, & H=300 \text{m},\\[2mm]
43, & H=1.3 \text{km}.
\end{cases}
\ee
Since the original ET vertical-displacement noise budget already lies comfortably below the detector sensitivity, this additional enhancement remains subdominant in both the Minimal Squashed Pyramid and Cliff-Extended configurations considered here. Therefore, introducing mildly non-coplanar arms does not lead to a significant degradation of the detector sensitivity while enabling direct access to the circular polarization of a cosmological gravitational-wave background.

\section{S3. Long-distance Seismic Correlations}

Measurements of underground seismic correlations over distances ranging from a few hundred meters up to about
$10\,{\rm km}$ show that long-distance coherence is mainly a low-frequency effect~\cite{Janssens:2024jln}. This can be understood from the seismic wavelength.
For two underground sites separated by $d\simeq 10\,{\rm km}$, coherent seismic motion requires
\[
\lambda(f)\gtrsim d,
\qquad \text{or equivalently} \qquad
f\lesssim \frac{v}{d}.
\]
Taking typical underground seismic velocities $v\sim 1$--$5\,{\rm km\,s^{-1}}$, this gives
\[
f_{\rm coh}\sim 0.1\text{--}0.5\,{\rm Hz}.
\]
Above $1\,{\rm Hz}$, the seismic wavelength is already shorter than the $10\,{\rm km}$ separation, and above a few Hz it is much shorter. Therefore, for a tilted underground arm separated from the standard underground ET arm by roughly $10\,{\rm km}$, any correlated seismic and Newtonian-noise contribution is expected to be relevant only at very low frequencies. In the ET band above a few Hz, the two underground environments can be treated as effectively decorrelated, up to narrow instrumental or site-dependent spectral lines.

In the detector pair considered here, \(D\) and \(D''\) share one horizontal arm, while the tilted arm of \(D''\)—both in the fully underground realization and in the cliff-extended configuration—is displaced by approximately \(10\,\mathrm{km}\) from the other horizontal arm of \(D\). Consequently, the seismic and Newtonian-noise contributions associated with the tilted arm are expected to be effectively uncorrelated with those of the standard underground arm over the frequency band relevant to our analysis.

In the minimal configuration with $H=300\,{\rm m}$, the tilted arm remains underground, so it is natural to use the ET underground noise budget as the appropriate benchmark. For the cliff-extended configuration, however, a significant part of the tilted arm may lie above ground. This can introduce additional low-frequency environmental noise sources, such as human activity, wind-induced motion, atmospheric-pressure fluctuations, and acoustic disturbances. We may schematically decompose the tilted-arm noise as
\[
n_{n''}(f)=n^{\rm ug}_{n''}(f)+\tilde n^{\rm og}_{n''}(f),
\]
where $n^{\rm ug}_{n''}$ denotes the underground ET-like contribution, while $\tilde n^{\rm og}_{n''}$ denotes the additional overground noise. Since the stochastic-background observable considered here relies on correlations with the underground coplanar ET arms, the overground contribution is expected to be largely uncorrelated with the underground noise,
\[
\left\langle
\tilde n^{\rm og}_{n''}(f)\,
n^{\rm ug\,*}_{n}(f')
\right\rangle
\ll
\left\langle
n^{\rm ug}_{n''}(f)\,
n^{\rm ug\,*}_{n}(f')
\right\rangle .
\]
We therefore neglect $\tilde n^{\rm og}_{n''}$ in the correlated-noise estimate above. This approximation is conservative for the correlation analysis, although a dedicated site-dependent environmental study would be required for a full instrumental noise budget of an above-ground extension.


\begin{thebibliography}{99}


\bibitem{Branchesi:2023}
M.~Branchesi \textit{et al.} (ET Science case),
%``Science with the Einstein Telescope,''
JCAP \textbf{07}, 068 (2023), % note: ET-D triangular 10 km reference sensitivity and science case.
%\bibitem{ET:2025xjr}
A.~Abac \textit{et al.} (ET Collaboration),
%``The Science of the Einstein Telescope,''
JCAP \textbf{03}, 081 (2026) .




\bibitem{Seto:2008sr}
  N.~Seto and A.~Taruya,
%``Polarization analysis of gravitational-wave backgrounds from the correlation signals of ground-based interferometers: Measuring a circular-polarization mode,''
Phys. Rev. D \textbf{77}, 103001 (2008)
%doi:10.1103/PhysRevD.77.103001
%[arXiv:0801.4185 [astro-ph]].
%99 citations counted in INSPIRE as of 01 Apr 2026


%\cite{Smith:2016jqs}
\bibitem{Smith:2016jqs}
T.~L.~Smith and R.~Caldwell,
%``Sensitivity to a Frequency-Dependent Circular Polarization in an Isotropic Stochastic Gravitational Wave Background,''
Phys. Rev. D \textbf{95}, no.4, 044036 (2017)
%doi:10.1103/PhysRevD.95.044036
%[arXiv:1609.05901 [gr-qc]].
%149 citations counted in INSPIRE as of 09 Apr 2026

%\cite{Duval:2025vfg}
\bibitem{Duval:2025vfg}
H.~Duval, C.~Badger and M.~Sakellariadou,
%``A battle of designs: triangular vs. L-shaped detectors and parity violation in the gravitational-wave background,''
[arXiv:2511.17422 [gr-qc]].
%0 citations counted in INSPIRE as of 06 Apr 2026



%\cite{Cook:2011hg}
\bibitem{Cook:2011hg}
J.~L.~Cook and L.~Sorbo,
%``Particle production during inflation and gravitational waves detectable by ground-based interferometers,''
Phys. Rev. D \textbf{85}, 023534 (2012)
%[erratum: Phys. Rev. D \textbf{86}, 069901 (2012)]
%doi:10.1103/PhysRevD.85.023534
%[arXiv:1109.0022 [astro-ph.CO]].
%381 citations counted in INSPIRE as of 03 Apr 2026
%\cite{Anber:2012du}
\bibitem{Anber:2012du}
M.~M.~Anber and L.~Sorbo,
%``Non-Gaussianities and chiral gravitational waves in natural steep inflation,''
Phys. Rev. D \textbf{85}, 123537 (2012)
%doi:10.1103/PhysRevD.85.123537
%[arXiv:1203.5849 [astro-ph.CO]].
%185 citations counted in INSPIRE as of 03 Apr 2026

%\cite{Dimastrogiovanni:2012ew}
%\bibitem{Dimastrogiovanni:2012ew}
\bibitem{Maleknejad:2011jw}
A.~Maleknejad and M.~M.~Sheikh-Jabbari,
%``Gauge-flation: Inflation From Non-Abelian Gauge Fields,''
Phys. Lett. B \textbf{723}, 224-228 (2013)
%doi:10.1016/j.physletb.2013.05.001
%[arXiv:1102.1513 [hep-ph]].
%313 citations counted in INSPIRE as of 03 Apr 2026
%\cite{Adshead:2013qp}
\bibitem{Adshead:2013qp}
P.~Adshead, E.~Martinec and M.~Wyman,
%``Gauge fields and inflation: Chiral gravitational waves, fluctuations, and the Lyth bound,''
Phys. Rev. D \textbf{88}, no.2, 021302 (2013)
%doi:10.1103/PhysRevD.88.021302
%[arXiv:1301.2598 [hep-th]].
%193 citations counted in INSPIRE as of 03 Apr 2026
\bibitem{Dimastrogiovanni:2012ew}
E.~Dimastrogiovanni and M.~Peloso,
%``Stability analysis of chromo-natural inflation and possible evasion of Lyth{\textquoteright}s bound,''
Phys. Rev. D \textbf{87}, no.10, 103501 (2013)
%doi:10.1103/PhysRevD.87.103501
%[arXiv:1212.5184 [astro-ph.CO]].
%168 citations counted in INSPIRE as of 03 Apr 2026
%\cite{Maleknejad:2016qjz}
\bibitem{Maleknejad:2016qjz}
A.~Maleknejad,
%``Axion Inflation with an SU(2) Gauge Field: Detectable Chiral Gravity Waves,''
JHEP \textbf{07}, 104 (2016)
%doi:10.1007/JHEP07(2016)104
%[arXiv:1604.03327 [hep-ph]].
%143 citations counted in INSPIRE as of 03 Apr 2026

%\cite{Alexander:2004us}
\bibitem{Alexander:2004us}
S.~H.~S.~Alexander, M.~E.~Peskin and M.~M.~Sheikh-Jabbari,
%``Leptogenesis from gravity waves in models of inflation,''
Phys. Rev. Lett. \textbf{96}, 081301 (2006)
%doi:10.1103/PhysRevLett.96.081301
%[arXiv:hep-th/0403069 [hep-th]].
%377 citations counted in INSPIRE as of 03 Apr 2026

%\cite{Satoh:2007gn}
\bibitem{Satoh:2007gn}
M.~Satoh, S.~Kanno and J.~Soda,
%``Circular Polarization of Primordial Gravitational Waves in String-inspired Inflationary Cosmology,''
Phys. Rev. D \textbf{77}, 023526 (2008)
%doi:10.1103/PhysRevD.77.023526
%[arXiv:0706.3585 [astro-ph]].
%219 citations counted in INSPIRE as of 03 Apr 2026

%\cite{Kharzeev:2007jp}
\bibitem{Kharzeev:2007jp}
D.~E.~Kharzeev, L.~D.~McLerran and H.~J.~Warringa,
%``The Effects of topological charge change in heavy ion collisions: 'Event by event P and CP violation',''
Nucl. Phys. A \textbf{803}, 227-253 (2008)
%doi:10.1016/j.nuclphysa.2008.02.298
%[arXiv:0711.0950 [hep-ph]].
%2040 citations counted in INSPIRE as of 06 Apr 2026

%\cite{Fukushima:2008xe}
\bibitem{Fukushima:2008xe}
K.~Fukushima, D.~E.~Kharzeev and H.~J.~Warringa,
%``The Chiral Magnetic Effect,''
Phys. Rev. D \textbf{78}, 074033 (2008)
%doi:10.1103/PhysRevD.78.074033
%[arXiv:0808.3382 [hep-ph]].
%2171 citations counted in INSPIRE as of 06 Apr 2026

%\cite{Kharzeev:2013ffa}
\bibitem{Kharzeev:2013ffa}
D.~E.~Kharzeev,
%``The Chiral Magnetic Effect and Anomaly-Induced Transport,''
Prog. Part. Nucl. Phys. \textbf{75}, 133-151 (2014)
%doi:10.1016/j.ppnp.2014.01.002
%[arXiv:1312.3348 [hep-ph]].
%496 citations counted in INSPIRE as of 06 Apr 2026

%\cite{Brandenburg:2023imm}
\bibitem{Brandenburg:2023imm}
A.~Brandenburg, E.~Clarke, T.~Kahniashvili, A.~J.~Long and G.~Sun,
%``Relic gravitational waves from the chiral plasma instability in the standard cosmological model,''
Phys. Rev. D \textbf{109}, no.4, 043534 (2024)
%doi:10.1103/PhysRevD.109.043534
%[arXiv:2307.09385 [astro-ph.CO]].
%5 citations counted in INSPIRE as of 06 Apr 2026

%\cite{Kamionkowski:1993fg}
\bibitem{Kamionkowski:1993fg}
M.~Kamionkowski, A.~Kosowsky and M.~S.~Turner,
%``Gravitational radiation from first order phase transitions,''
Phys. Rev. D \textbf{49}, 2837-2851 (1994)
%doi:10.1103/PhysRevD.49.2837
%[arXiv:astro-ph/9310044 [astro-ph]].
%1027 citations counted in INSPIRE as of 06 Apr 2026



%\cite{Witten:1984rs}
\bibitem{Witten:1984rs}
E.~Witten,
%``Cosmic Separation of Phases,''
Phys. Rev. D \textbf{30}, 272-285 (1984)
%doi:10.1103/PhysRevD.30.272
%3722 citations counted in INSPIRE as of 06 Apr 2026


%\cite{Brandenburg:2021aln}
\bibitem{Brandenburg:2021aln}
A.~Brandenburg, Y.~He, T.~Kahniashvili, M.~Rheinhardt and J.~Schober,
%``Relic gravitational waves from the chiral magnetic effect,''
Astrophys. J. \textbf{911}, no.2, 110 (2021)
%doi:10.3847/1538-4357/abe4d7
%[arXiv:2101.08178 [astro-ph.CO]].
%36 citations counted in INSPIRE as of 06 Apr 2026

%\cite{Adshead:2015pva}
\bibitem{Adshead:2015pva}
P.~Adshead, J.~T.~Giblin, T.~R.~Scully and E.~I.~Sfakianakis,
%``Gauge-preheating and the end of axion inflation,''
JCAP \textbf{12}, 034 (2015)
%doi:10.1088/1475-7516/2015/12/034
%[arXiv:1502.06506 [astro-ph.CO]].
%221 citations counted in INSPIRE as of 07 Apr 2026

%\cite{Machado:2018nqk}
\bibitem{Machado:2018nqk}
C.~S.~Machado, W.~Ratzinger, P.~Schwaller and B.~A.~Stefanek,
%``Audible Axions,''
JHEP \textbf{01}, 053 (2019)
%doi:10.1007/JHEP01(2019)053
%[arXiv:1811.01950 [hep-ph]].
%100 citations counted in INSPIRE as of 07 Apr 2026

%\cite{Maleknejad:2012fw}
\bibitem{Maleknejad:2012fw}
A.~Maleknejad, M.~M.~Sheikh-Jabbari and J.~Soda,
%``Gauge Fields and Inflation,''
Phys. Rept. \textbf{528}, 161-261 (2013)
%doi:10.1016/j.physrep.2013.03.003
%[arXiv:1212.2921 [hep-th]].
%381 citations counted in INSPIRE as of 06 Apr 2026

%\cite{Komatsu:2022nvu}
\bibitem{Komatsu:2022nvu}
E.~Komatsu,
%``New physics from the polarized light of the cosmic microwave background,''
Nature Rev. Phys. \textbf{4}, no.7, 452-469 (2022)
%doi:10.1038/s42254-022-00452-4
%[arXiv:2202.13919 [astro-ph.CO]].
%196 citations counted in INSPIRE as of 06 Apr 2026
    

%\cite{LISACosmologyWorkingGroup:2022jok}
\bibitem{LISACosmologyWorkingGroup:2022jok}
P.~Auclair \textit{et al.} [LISA Cosmology Working Group],
%``Cosmology with the Laser Interferometer Space Antenna,''
Living Rev. Rel. \textbf{26}, no.1, 5 (2023)
%doi:10.1007/s41114-023-00045-2
%[arXiv:2204.05434 [astro-ph.CO]].
%576 citations counted in INSPIRE as of 07 Apr 2026

%\cite{Maleknejad:2025clz}
\bibitem{Maleknejad:2025clz}
A.~Maleknejad,
%``When Geometry Radiates Review: Gravitational Waves in Theory, Cosmology, and Observation,''
[arXiv:2512.21328 [gr-qc]].
%1 citations counted in INSPIRE as of 07 Apr 2026

%\cite{Minami:2020odp}
\bibitem{Minami:2020odp}
Y.~Minami and E.~Komatsu,
%``New Extraction of the Cosmic Birefringence from the Planck 2018 Polarization Data,''
Phys. Rev. Lett. \textbf{125}, no.22, 221301 (2020)
%doi:10.1103/PhysRevLett.125.221301
%[arXiv:2011.11254 [astro-ph.CO]].
%308 citations counted in INSPIRE as of 07 Apr 2026

\bibitem{SetoTaruya:2007}
N.~Seto and A.~Taruya,
%``Measuring a Parity Violation Signature in the Early Universe via Ground-based Laser Interferometers,''
Phys. Rev. Lett. \textbf{99}, 121101 (2007)
%doi:10.1103/PhysRevLett.99.121101
%[arXiv:0707.0535 [astro-ph]].
%127 citations counted in INSPIRE as of 06 Apr 2026
%\bibitem{Seto:2008sr}
 


 \begin{comment}
\bibitem{KingdomTower}
A.~Smith and G.~Gill,
%``Case Study: Kingdom Tower, Jeddah,''
in {\it Proceedings of the CTBUH 2015 New York Conference}
(Council on Tall Buildings and Urban Habitat, New York, 2015).  
 \end{comment}
 
\bibitem{Danilishin:2019dxq}
S.~L.~Danilishin, F.~Ya.~Khalili, and H.~Miao,
%``Advanced quantum techniques for future gravitational-wave detectors,''
Living Rev.\ Relativ.\ \textbf{22}, 2 (2019).


\bibitem{LartauxCapocasa:2025}
A.~Lartaux and E.~Capocasa,
%\textit{Squeezing activities in France for Einstein Telescope},
Talk presented at the Réunion ET France 2025 (IJCLab), 2025 %,\\
%\url{https://indico.in2p3.fr/event/37047/contributions/162513/attachments/96767/148770/2025_10_08_ET_France_Squeezing.pdf}.

\bibitem{SM}
See Supplemental Material for further details on the main known noise sources, vertical suspension noise in the tilted-arm configuration, and supporting figures.

%\cite{Aggarwal:2025noe}
\bibitem{Aggarwal:2025noe}
N.~Aggarwal  \textit{et al.}
%``Challenges and opportunities of gravitational-wave searches above 10 kHz,''
Living Rev. Rel. \textbf{28}, no.1, 10 (2025)
%doi:10.1007/s41114-025-00060-5
%[arXiv:2501.11723 [gr-qc]].
%99 citations counted in INSPIRE as of 06 Apr 2026
%\cite{Kharzeev:2025lyu}
\bibitem{Kharzeev:2025lyu}
D.~E.~Kharzeev, A.~Maleknejad and S.~Shalamberidze,
%``QuGrav: Bringing gravitational waves to light with qumodes,''
Phys. Rev. Res. \textbf{8}, no.1, 013140 (2026)
%doi:10.1103/nbzp-1yn7
%[arXiv:2506.09459 [gr-qc]].
%5 citations counted in INSPIRE as of 06 Apr 2026

%\cite{Evans:2021gyd}
\bibitem{Evans:2021gyd}
M.~Evans \textit{et al.},
%``A Horizon Study for Cosmic Explorer: Science, Observatories, and Community,''
[arXiv:2109.09882 [astro-ph.IM]].
%803 citations counted in INSPIRE as of 07 Apr 2026



%\cite{Thrane:2013oya}
\bibitem{Thrane:2013oya}
E.~Thrane and J.~D.~Romano,
%``Sensitivity curves for searches for gravitational-wave backgrounds,''
Phys. Rev. D \textbf{88}, no.12, 124032 (2013)
%doi:10.1103/PhysRevD.88.124032
%[arXiv:1310.5300 [astro-ph.IM]].
%647 citations counted in INSPIRE as of 06 Apr 2026


\bibitem{Bertocco:2024}
A.~Bertocco, et. al. 
%``New Generation of Superattenuator for Einstein Telescope: preliminary studies,''
\emph{Class. Quantum Grav.} \textbf{41}, no.~11, 117004 (2024) %, doi:10.1088/1361-6382/ad407e.

%\cite{Janssens:2024jln}
\bibitem{Janssens:2024jln}
K.~Janssens, \textit{et al.}
%``Correlated 0.01{\textendash}40~Hz seismic and Newtonian noise and its impact on future gravitational-wave detectors,''
Phys. Rev. D \textbf{109}, no.10, 102002 (2024)
%doi:10.1103/PhysRevD.109.102002
%[arXiv:2402.17320 [gr-qc]].
%18 citations counted in INSPIRE as of 06 May 2026

\end{thebibliography}
\end{document}